\documentclass[aps,prl,reprint,superscriptaddress]{revtex4-1}

\usepackage{graphicx}
\usepackage{bm}
\usepackage{amssymb}
\usepackage{amsmath}
\usepackage{epsfig}
\usepackage{rotating}
\usepackage{verbatim}
\usepackage{wrapfig}
\usepackage{float}
\usepackage{subfigure}
\usepackage[export]{adjustbox}
\usepackage[skip=2pt,font=small]{caption}
\usepackage{ftnxtra}
\usepackage{physics}
\usepackage{graphicx}
\usepackage{dcolumn}
\usepackage{bm}

\providecommand\bcdot{\boldsymbol{\cdot}}
\newcommand\bnabla{\boldsymbol{\nabla}}

\begin{document}


\title{A continuous pathway between the elasto-inertial and elastic turbulent states in viscoelastic channel flow}
\author{Mohammad Khalid}
\affiliation{Department of Chemical Engineering, Indian Institute of Technology, Kanpur 208016, India}

\author{V. Shankar}
\email{vshankar@iitk.ac.in}
\affiliation{Department of Chemical Engineering, Indian Institute of Technology, Kanpur 208016, India}

\author{Ganesh Subramanian}
\email{sganesh@jncasr.ac.in}
\affiliation{Engineering Mechanics Unit, Jawaharlal Nehru Center for Advanced Scientific Research, Bangalore 560084, India}


%


\date{\today}

\begin{abstract}
We show that viscoelastic plane Poiseuille flow becomes linearly unstable in the absence of inertia, in the limit of high elasticities, for ultra-dilute polymer solutions. While inertialess elastic instabilities have been predicted for curvilinear shear flows, this is the first ever report of a purely elastic linear instability in a rectilinear shear flow. The novel instability continues upto a Reynolds number ($Re$) of $O(1000)$, corresponding to the recently identified elasto-inertial turbulent state believed to underlie the maximum-drag-reduced regime. Thus, for highly elastic ultra-dilute polymer solutions, a single linearly unstable modal branch may underlie transition to elastic turbulence at zero $Re$, and to elasto-inertial turbulence at moderate $Re$, implying the existence of continuous pathways connecting the turbulent states to each other, and to the laminar base state. 

\end{abstract}

\pacs{}

\maketitle


Dilute polymer solutions undergo two different transitions to novel turbulent states, both driven by viscoelasticity, and thence, fundamentally distinct from the now well-understood Newtonian transition \citep{kerswell_2005,eckhardt_etal_2007,avila2011}.  
Rectilinear shearing flows of sufficiently elastic polymer solutions (e.g., pipe and channel flow) transition from the laminar state at Reynolds numbers substantially lower than the typical Newtonian threshold, the ensuing flow state dubbed `elastoinertial turbulence' (EIT) to emphasize the importance of both elastic and inertial effects underlying the turbulent dynamics \citep{Samanta2013,Choueiri2018,Bidhan2018,choueiri2021experimental,lopez_choueiri_hof_2019,Shekar2019,Shekar_2020}. The EIT state is dominated by spanwise oriented 2D structures \citep{Sid_2018_PRF,lopez_choueiri_hof_2019} in sharp contrast to the Newtonian scenario \citep{eckhardt_etal_2007}, and for plane-Poiseuille flow, has recently been shown  \citep{page2020exact,dubief2020coherent} to be connected subcritically to a linear center-mode instability  \citep{Piyush_2018,chaudharyetal_2021,khalid2020centermode}; recent pipe-flow experiments have, in fact, found remarkable agreement between the observed structures at EIT onset \citep{choueiri2021experimental} and the center-mode eigenfunction \citep{Piyush_2018,chaudharyetal_2021}.
On the other hand, curvilinear shearing flows of dilute polymer solutions transition to `elastic turbulence' (ET), with inertial effects being negligibly small \citep{Groisman2000,Groisman2001,Steinberg_AFM}. This latter transition has been known for a longer time, and its origin may be traced to a hoop-stress-driven linear instability in such flows \citep{larson_shaqfeh_muller_1990,McKinley1991,oztekin_brown_1993,joo_shaqfeh_1994,Pakdel_McKinley,shaqfeh1996,Schiamberg_2006,Jun_Steinberg_2011}; the eventual disorderly ET state that arises has been well characterized experimentally \citep{Groisman2000,Groisman2001,Jun_Steinberg_2011},  and to a limited extent, theoretically as well
\citep{fouxon_lebedev}.


Unlike their curvilinear counterparts, inertialess rectilinear shear flows of dilute polymer solutions (where shear thinning effects are insignificant), have  hitherto been regarded as linearly stable \citep{Hodenn1977,larson1992,wilson1999}. Transition in these flows has been proposed to occur via a subcritical mechanism, but one that nevertheless involves a hoop stress that now arises at a nonlinear order due to the curvature of the perturbed streamlines 
\citep{bertola_saarloos2003,meulenbroek_sarloos2004,morozov_saarloos2005,morozov_saarloos2007,morozov_saarloos2019}. There is some experimental evidence of an inertialess finite-amplitude transition \citep{bertola_saarloos2003,Pan_2012_PRL,Arratia_PRL_2017,Arratia_PRL_2019}, to what might be an ET state similar to that observed for curvilinear flows; the absence of a linear instability has also motivated examination of  non-modal growth mechanisms, both in the absence \citep{jovanovic_kumar_2010,jovanovic_kumar_2011} and presence \citep{Zaki2013} of inertia. The aforementioned EIT and ET states might seem unrelated at first sight, on account of fluid 
inertia playing a fundamental role in the former while being irrelevant in the latter. It has nevertheless been speculated \citep{Samanta2013,page2020exact,Arratia_PRL_2019,Steinberg_AFM,choueiri2021experimental} that the two states may be linked, although there exist no concrete hypotheses or proposals in this regard.

In sharp contrast to the prevalent view above, in this Letter, we show that: (i) pressure-driven
viscoelastic channel flow is linearly unstable even in the absence of inertia, making this the first ever report of a purely elastic linear instability not dependent on base-state streamline curvature; (2) the  instability smoothly continues, with increasing $Re$, to 
the aforementioned elasto-inertial linear instability \citep{Piyush_2018,khalid2020centermode}. Since the latter instability
has been shown to subcritically continue to nonlinear elastoinertial coherent structures \citep{page2020exact,dubief2020coherent}, the implication is the existence of a continuous pathway (the underlying unstable modal branch) connecting the EIT and ET states, one that might provide a template for nonlinear coherent structures acting as possible bridges between these states, thereby forming the framework for a dynamical systems based interpretation of turbulence outside the Newtonian realm. 


\begin{figure}
\begin{center}
\includegraphics[scale=0.3]{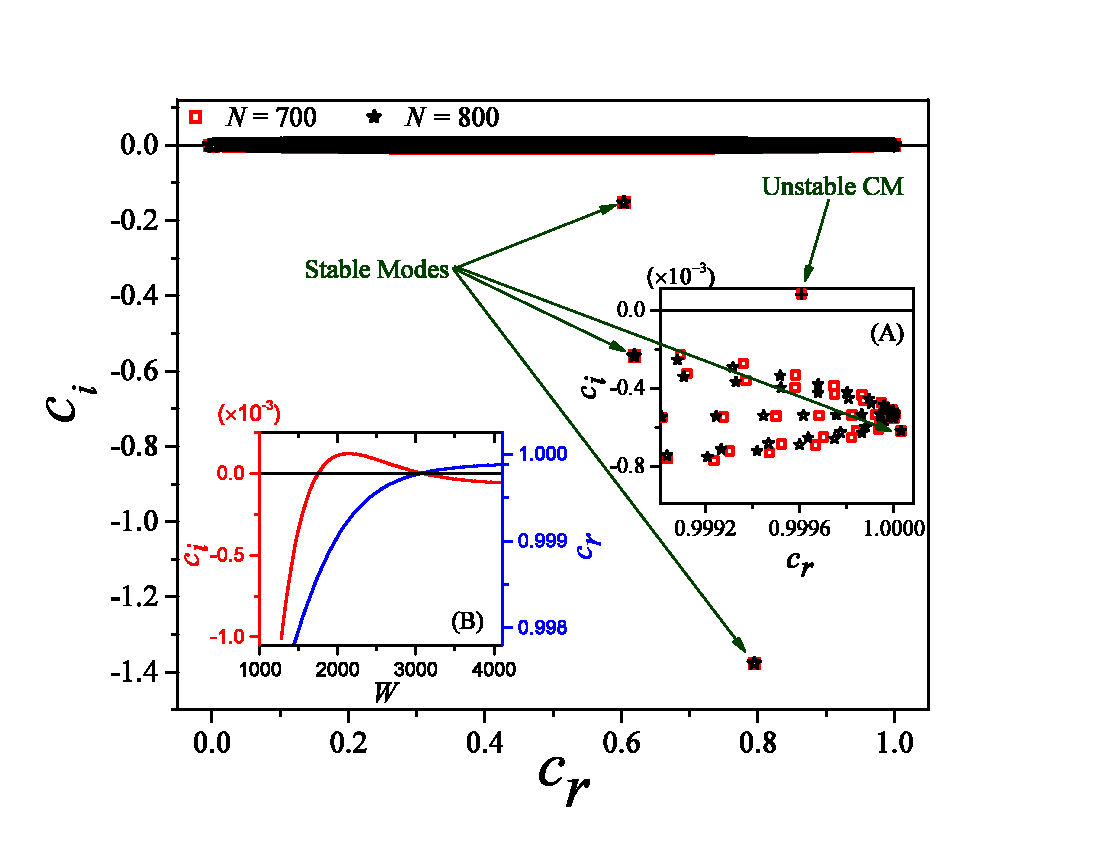}
\end{center}
\caption{\label{fig:spectrum} The spectrum for inertialess plane Poiseuille flow of an Oldroyd-B fluid shows five discrete modes for $\beta = 0.997$, $k = 0.75$, $W = 2500$, with three in the main figure and the remaining two visible in inset~(A); $N$ is the number of Chebyshev polynomials in the spectral expansion.  Inset~A shows an enlarged view of the region near the unstable center mode (labelled CM). Inset~B shows the variation of $c_r$ and $c_i$ with $W$ for $\beta = 0.997$ and $k = 0.75$.}
\end{figure}

We consider pressure-driven flow of an incompressible viscoelastic fluid in a channel of width $2H$.
The governing mass and momentum equations \cite{birdvol1,larson1988constitutive},  in dimensionless form, are 
\begin{equation}
\bnabla \bcdot \mathbf{u}=0, \, 
Re \, \Big(\frac{\partial \mathbf{u}}{\partial t}+(\mathbf{u}\bcdot\bnabla) \, \mathbf{u} \Big)=-\bnabla {p}+ {\beta} \,\nabla^2 \, \mathbf{u} +\bnabla \bcdot \boldsymbol{\tau} ,
\label{eq:fluid-momentum}
\end{equation}
where $\mathbf{u}$, $p$, and $\boldsymbol{\tau}$, are the velocity, pressure, and the polymer stress fields respectively. Here, lengths are nondimensionalized using the channel half-width $H$, velocities using the base-flow maximum $U_{max}$, and $\boldsymbol{\tau}$ is assumed to be governed by the Oldroyd-B constitutive relation
\begin{align}
\boldsymbol{\tau}+{W} \,\Big(\frac{\partial \boldsymbol{\tau}}{\partial t}+(\mathbf{u}\bcdot\bnabla )\boldsymbol{\tau}-&(\bnabla \mathbf{u})^{T} \bcdot \boldsymbol{\tau}- \,\boldsymbol{\tau} \bcdot(\bnabla \mathbf{u})\Big) \notag \\
&=\Big( 1-\beta \Big)\Big(\bnabla \mathbf{u}+\bnabla \mathbf{u}^T \Big).
\label{eq:stress constitutive eq}
\end{align}
The relevant dimensionless groups are the 
Reynolds number $Re = \rho U_{max} H/\eta$ ($\rho$ and $\eta$ being the density and viscosity of the polymer solution, respectively), the ratio of solvent to solution viscosities $\beta$, and the Weissenberg number $W = \lambda U_{max}/H$, where $\lambda$ is the polymer relaxation time; below, we also use the elasticity number $E = W/Re$ in lieu of $W$. 
The Oldroyd-B model regards the polymer molecules as non-interacting Hookean dumbbells, resulting in a shear-rate-independent viscosity and first normal stress coefficient in viscometric flows, 
and has been  successfully used to predict linear and nonlinear instabilities in
rectilinear \citep{sureshkumar1995linear, morozov_saarloos2005,Piyush_2018}, and both
curvilinear viscometric \citep{larson_shaqfeh_muller_1990,McKinley1991,joo_shaqfeh_1994}, and non-viscometric \cite{Poole2007} flows.  For plane Poiseuille flow, whose stability is examined here, the laminar state is given by $U(z) = (1-z^2)$, with a first normal stress difference $T_{xx} - T_{zz} = 8(1-\beta)W^2 z^2$. We analyze the temporal stability of this state by imposing infinitesimal 2D perturbations
(justified by the Squire's theorem  \citep{BISTAGNINO2007})
 of the form $f'(x,z,t) = \tilde{f}(z) \exp[ik(x-ct)]$, where $f'(x,z,t)$ is the perturbation to any dynamical variable, $\tilde{f}(z)$ is the eigenfunction, $k$ is the streamwise wavenumber, and $c = c_r + i c_i$ is the complex wavespeed with the flow being temporally unstable when $c_i > 0$.  Linearization of the governing equations results in a generalized eigenvalue problem \citep{sureshkumar1995linear,khalid2020centermode} which is solved using spectral collocation and shooting methods, both of which have been extensively benchmarked in our earlier efforts \citep{chaudhary_etal_2019,chaudharyetal_2021,khalid2020centermode}. 

\begin{figure}
\includegraphics[scale=0.25]{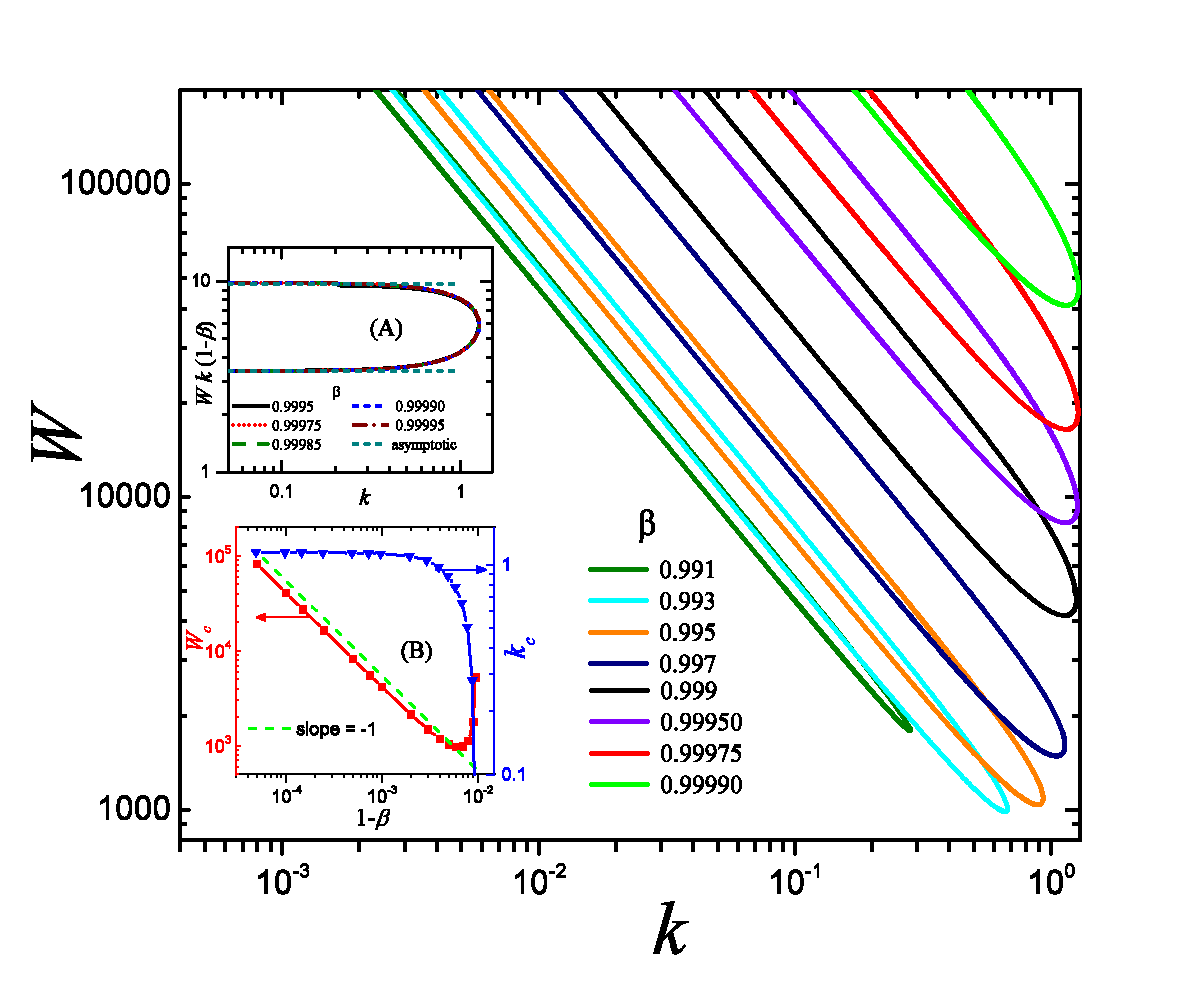}
\caption{\label{neutralcurves} Neutral curves in the $W$--$k$ plane for different $\beta$'s in the creeping-flow limit. Inset~A shows the collapse of neutral curves for $\beta \rightarrow 1$ when plotted as $Wk(1-\beta)$ vs $k$, and the results obtained from the reduced equations in that limit.
Inset~B shows the variation of the critical Weissenberg number $W_c$ and wavenumber $k_c$ with $(1-\beta)$. }
\end{figure}

%

We first consider the creeping-flow limit ($Re = 0$) where the relevant dimensionless groups are $W$ and $\beta$. In the absence of the solvent ($\beta = 0$; the upper-convected Maxwell (UCM) limit), the inertialess plane Poiseuille  eigenspectrum is known \citep{wilson1999} to contain six discrete modes for $W, k \sim O(1)$, in addition to the continuous spectrum (CS). As $W$ is increased for a fixed $k \sim O(1)$ (or, as $k$ is increased at a fixed $W \sim O(1)$), two of the modes merge into the CS, while two transition into `wall modes'. The remaining two transition into `center modes' with phase speeds approaching the  base-flow maximum. Regardless of $W$ and $k$, however, plane Poiseuille flow of a UCM fluid always remains stable \citep{wilson1999}. For
$\beta \neq 0$, the number of discrete modes depends on  $\beta$, $k$ and $W$. For ultra-dilute solutions with  $\beta > 0.99$, there are two center modes out of a total of five discrete modes (Fig.~\ref{fig:spectrum}).  Interestingly, for $\beta > 0.9905$, 
at sufficiently high $W \sim 2500$, one of the two center modes becomes unstable (inset~A of Fig.~\ref{fig:spectrum}). The instability runs counter to the prevailing view \citep{wilson1999}, and 
the unexpected destabilizing role of the solvent viscosity is similar to the recently discovered elastoinertial center-mode instability \citep{Piyush_2018,chaudharyetal_2021,khalid2020centermode}. 
The regime $W \sim O(10^3)$, $\beta > 0.99$ corresponds to highly-elastic ultra-dilute polymer solutions (with concentrations around 1\% of the overlap value), one that was not explored in earlier theoretical work \citep{wilson1999}, but 
has  recently been shown to be experimentally accessible in microscale flows \citep{Varshney_2018_creeping,jha2020universal}. 

Neutral curves demarcate unstable `tongues' in the $W$-$k$ plane (Fig.~\ref{neutralcurves}), with the tongues ceasing to exist beyond a critical $k$ and below a threshold $W$. However, the instability continues to exist at arbitrarily small $k$, with $W \sim 1/k$ for $k \ll 1$ along both the branches of the tongue. 
A plot of $W_c$ (the critical $W$ corresponding to the minimum along a neutral curve) with $(1-\beta)$ (inset~B of Fig.~\ref{neutralcurves}) shows that the lowest $W_c$ is $\sim 973.8$ for $\beta = 0.994$, and that $W_c \propto (1-\beta)^{-1}$, $k_c \sim O(1)$ for $\beta \rightarrow 1$; 
 expectedly, $W(1-\beta)k$ is the threshold parameter for $k, (1-\beta) \ll 1$; see inset~A of
 Fig.~\ref{neutralcurves} and \footnote{The lower and upper threshold values of $Wk(1-\beta)$ for the instability are captured very well by a numerical solution of the reduced set of governing equations in this limit}.
The instability ceases to exist below $\beta = 0.9905$.

\begin{figure}
\centering
   \subfigure[\, Contours of ${v}_x$ for $W(1-\beta) = 43$ (A) and $80$ (B)]{
    \includegraphics[scale=0.25]{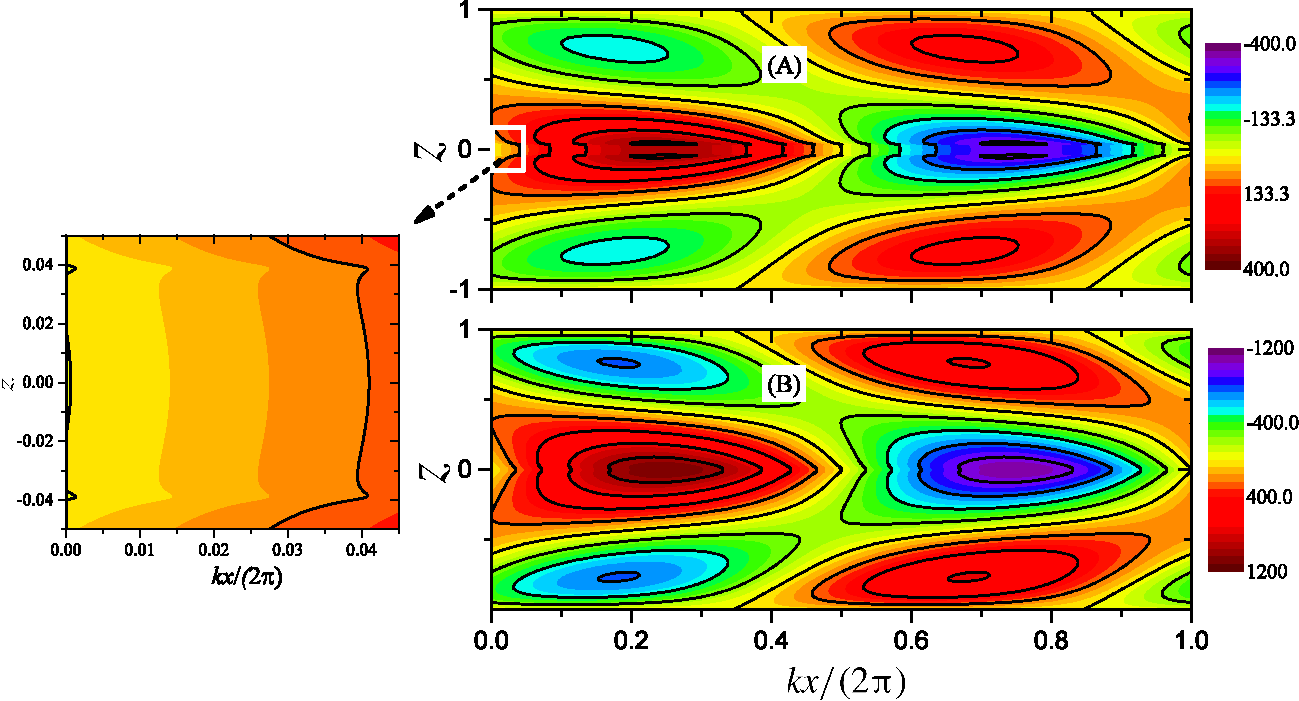}
    \label{subfig2a}
     \quad
  } \subfigure[\, $\tilde{\tau}_{xx}$ and $\tilde{\tau}_{xz}$ for $W(1-\beta) = 43$ (A) and $80$ (B)]{
    \includegraphics[scale=0.25]{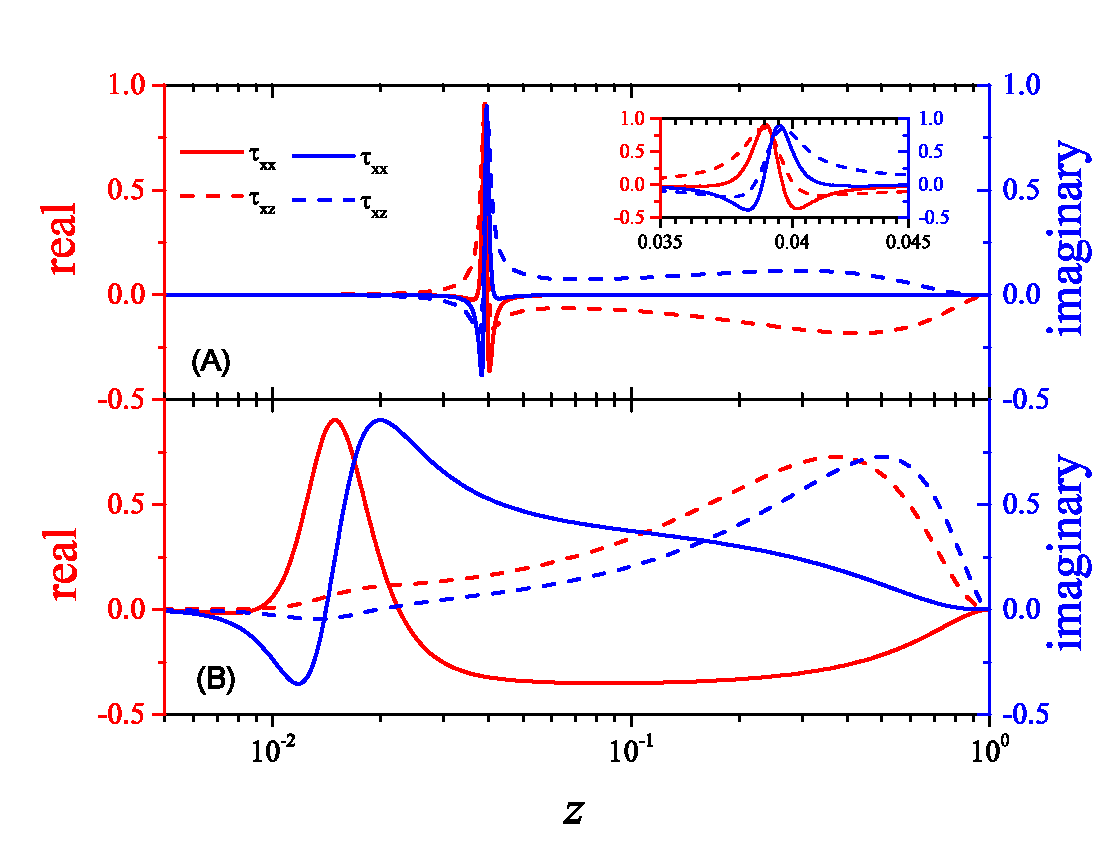}
    \label{subfig2b}
  }
\caption{Eigenfunctions for the center mode for $W (1-\beta) = 43$, $80$ and $k = 0.08$,  corresponding to neighborhood of the lower threshold and the center of the unstable region in the collapsed neutral curves shown in  inset~(A) of Fig.~\ref{neutralcurves}. The constant-amplitude
contours of the streamwise velocity are shown in subfigure~(a),  while  the eigenfunctions  for  $\tau_{xz}$ and $\tau_{xx}$ are
shown in subfigure~(b).  The expanded region near the centerline for panel~(A) in subfigure~(a) is also shown.}
\label{eigfuns}
\end{figure}

Figure~\ref{subfig2a} shows contour plots for the streamwise velocity field alongside the polymeric stress eigenfunctions in Fig.~\ref{subfig2b} for $W(1-\beta)$'s near the lower threshold  and close to the maximum growth rate (inset~A of Fig.~\ref{neutralcurves}). The instability arises due to stretched polymers being rotated away from flow-alignment by the perturbation shear, as they are swept past by the base-state parabolic flow. The differential rate of convection becomes small close to the centerline, owing to the phase speed of the eigenmode closely approaching the base-state maximum. As a result, the time available for the perturbation-shear-induced rotation (of the stretched polymers) increases, and the resulting accumulation of perturbation elastic shear stress ($\tau_{xz}$) drives a reinforcing flow, leading to exponential growth.
Close to neutrality, the stress ($\tau_{xx}$ and $\tau_{xz}$) eigenfunctions in particular (panel~A of Fig.~\ref{subfig2b}), are seen to develop singular features in an $O(1/W)$ elastic critical layer, where the phase speed equals the local laminar velocity, and this leads to additional rolls close to the centerline. The complicated eigenfunction structure near the lower threshold, as also evident from the cusps close to the centerline (see magnified view in Fig.~\ref{subfig2a}), arises from competing influences of $\tau_{xz}$ (destabilizing, as mentioned above) and $\tau_{xx}$ (stabilizing). Note that the existence of the singular elastic critical layer above, while emphasizing a hoop-stress-independent pathway to instability, also contrasts with the Newtonian scenario where the unstable inflectional mode approaches a regular neutral mode at the critical $Re$ \cite{Drazinreid}.

\begin{figure}
\includegraphics[scale=0.3]{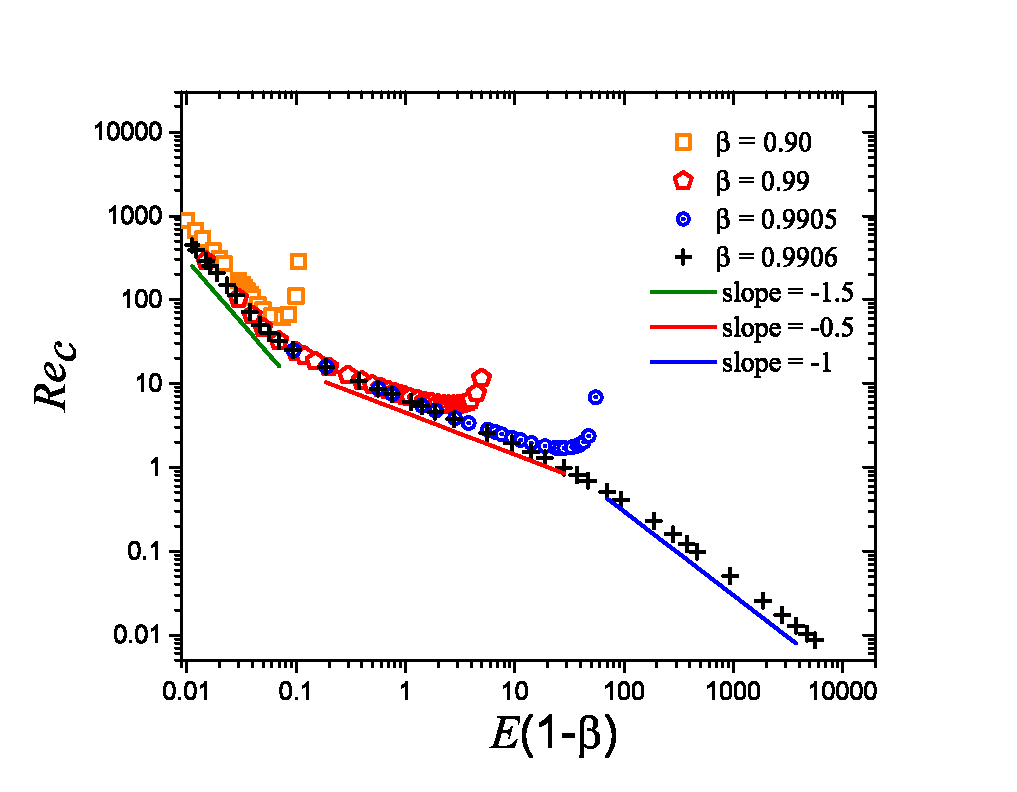}
\caption{\label{fig:RevsEoneminusbita} The variation of critical Reynolds number $Re_c$ with $E(1-\beta)$ for different $\beta$'s for $\beta \geq 0.9$. For $\beta$ up to $0.9905$, $Re_c$ decreases as $Re_c \propto (E(1-\beta))^{-3/2}$ for small $E(1-\beta)$, but increases sharply beyond a critical $E(1-\beta)$. For $\beta > 0.9905$, while $Re_c$ initially decreases as $Re_c \propto (E(1-\beta))^{-3/2}$, beyond a critical $E(1-\beta)$, there is a crossover to the scaling $Re_c \propto (E(1-\beta))^{-1}$, with the instability persisting down to the creeping-flow limit. For $0.98<\beta<0.9905$, there is an intermediate scaling regime with $Re_c \propto (E(1-\beta))^{-1/2}$.}
\end{figure}

The results above highlight a hitherto unknown center-mode instability in inertialess channel flow of an  Oldroyd-B fluid for $\beta > 0.9905$, whereas in earlier efforts \citep{Piyush_2018,chaudharyetal_2021,khalid2020centermode} we have
identified a center-mode instability  for $Re \sim O(100)$ and $\beta \sim 0.9$, both for viscoelastic channel and pipe flows, and that may underlie the EIT state  in these geometries \citep{Samanta2013,Choueiri2018,Srinivas-Kumaran2017}. The question arises, naturally, as to whether there is any connection between the elastic center mode identified here, and the elasto-inertial one identified in Refs.~\citep{Piyush_2018,chaudharyetal_2021,khalid2020centermode}. Figure~\ref{fig:RevsEoneminusbita} shows the critical Reynolds number $Re_c$ as a function of $E(1-\beta)$ for different $\beta$, and confirms such a connection. For $\beta \leq 0.9905$, the  elasto-inertial instability \citep{Piyush_2018} exhibits the  scaling $Re_c \propto (E(1-\beta))^{-3/2}$ for $E(1-\beta) \ll 1$, reflecting the simultaneous importance of  inertia, elasticity and viscous effects in a  boundary layer near the channel centerline \citep{Piyush_2018}, with $Re_c$ increasing sharply beyond a threshold $E(1-\beta)$.
 In stark contrast, for $\beta > 0.9905$, while $Re_c$ initially decreases as $(E(1-\beta))^{-3/2}$, there is a crossover to $Re \propto (E(1-\beta))^{-1}$ for higher $E(1-\beta)$, with this scaling persisting down to arbitrarily small $Re$. The $Re \propto E^{-1}$ scaling translates to an independence with respect to fluid inertia, and indeed corresponds to the creeping-flow instability identified above, with $W(1-\beta)$ being the threshold parameter.
 While a connection between EIT and ET states has been conjectured  \citep{Samanta2013,page2020exact,Steinberg_AFM}, and the possibility of a common instability underlying these states speculated upon \citep{Arratia_PRL_2019}, ours is the first explicit demonstration of the same.  Interestingly, as seen in Fig.~\ref{fig:RevsEoneminusbita},  an intermediate scaling regime with $Re_c \propto (E(1-\beta))^{-1/2}$ begins to emerge for $\beta  > 0.99$. This scaling implies that the flow velocity $V \propto \frac{1}{(1-\beta)} V_{shear}$, $V_{shear} = \sqrt{\frac{\eta (1-\beta)} {\lambda \rho}}$ being the elastic shear wave speed. The shear wave signature in this intermediate regime of relatively low $Re$ and high $E$ could be 
relevant to the observation of `waves' in experiments involving the flow of highly elastic polymer solutions \citep{Varshney_2018_creeping,Varshney_2019_Alfen,jha2020universal}.  Continuity considerations imply that the $\beta$'s where $Re_c$ increases sharply, and those where $Re_c$ transitions to an $E^{-1}$ scaling for $E$'s beyond the intermediate regime, are likely separated by a critical $\beta$ where, remarkably, the aforesaid shear-wave scaling must persist in the limit $Re \rightarrow 0$, $E \rightarrow \infty$. The relevance of a shear wave at low $Re$, in the flow of highly dilute solutions,  is intriguing especially 
in light of the fact that the class of shear waves originally analyzed in plane shear flow of a UCM fluid for $Re \ll 1$ \citep{GORODTSOV1967} are heavily damped for small but finite $\beta$ \citep{chaudhary_etal_2019}. 
 
Traditional research in Newtonian turbulence has been rooted in a statistical description aiming to explain, for instance, the subtle nature of small-scale universality at high $Re$, in a homogeneous isotropic setting \citep{frisch_1995,sreenivasan_antonia,sreenivasan1999}. The approach at moderate $Re$, especially from the perspective of understanding transition, has, however, undergone a paradigm shift with the advent of a dynamical systems perspective \citep{barkley2016,budanur_etal_2017}. For the canonical rectilinear shearing flows, this latter approach is based on the dynamics of so-called exact coherent structures that account for both an ambient shear and the presence of boundaries \citep{Fabian_PRL_1998,waleffe_2001,wedin_kerswell_2004,Cvitanovic_2016,budanur_etal_2017}. In contrast, theoretical efforts aimed at understanding the effect of elasticity on, for instance, the energy 
cascade in Newtonian turbulence, mirror the statistical approach above \citep{balkovksy_fouxon_lebedev,balkovsky_fouxon_lebedev_2001}. For elastic turbulence, in particular, such efforts have predicted a rapid decay, at least as fast as $k^{-3}$ ($k$ being the wavenumber), of the kinetic energy spectrum pointing to the spatially smooth character of small-scale elastic turbulent fluctuations, an aspect that has found experimental confirmation \citep{Groisman2000,Groisman2001};  there remains a dearth of information on the structural front, however. Recent experimental efforts, although still limited to curvilinear geometries, have begun to explore this latter aspect \citep{Jun_Steinberg_2011,Steinberg_AFM}; a recent theoretical work  has also explored the near-wall boundary layer structure in the ET state \citep{belan_chernykh_lebedev_2018}. 

Our discovery of a linear instability that spans the ET and EIT regimes (Fig.~\ref{fig:RevsEoneminusbita}) helps significantly expand the above picture, opening up multiple avenues for future research.  A linear elastic instability, with a physical origin genuinely different from the hoop-stress-based pathway in curvilinear shearing flows, offers a  template for novel nonlinear elastic coherent structures  that could shed light on the large-scale dynamics of the ET state; this would complement the aforementioned statistical approach tailored to the smallest scales. Such an approach also offers an alternative to prevailing efforts that analyze the elastic transition in rectilinear shearing flows, based on a bifurcation-from-infinity perspective \citep{meulenbroek_sarloos2004,morozov_saarloos2005,morozov_saarloos2007}, by expanding about a linearly stable eigenmode.  The physical basis of the nonlinear expansion is intimately tied to the hoop-stress-based mechanism that leads to a linear instability in the curvilinear geometries. It is also worth noting that these nonlinear analyses were restricted to the $\beta \rightarrow 0$ limit \citep{morozov_saarloos2005,morozov_saarloos2019}.
As evident from Fig.~\ref{fig:spectrum}, and the description of the eigenspectrum above (the stability of the UCM limit in particular), the elastic eigenspectrum is sensitively dependent on $\beta$, and the validity of a $\beta \rightarrow 0$ analysis, for the experimentally relevant case of dilute solutions with $\beta \rightarrow 1$, is not obvious. 
 An expansion about, or a numerical continuation from, the unstable eigenmode reported here offers access to a much
larger region in the viscoelastic parameter space comprising $W$, $Re$, and $\beta$. The recent finding  of a connection between an EIT structure \citep{page2020exact,dubief2020coherent} and the underlying linear eigenfunction \citep{Piyush_2018} points to the likely success of a numerical continuation approach. 
Finally, the nonlinear coherent structures might be relevant to a wider range of polymer concentrations and Weissenberg numbers than the restricted range ($\beta > 0.9905$, $W \sim O(1000)$) corresponding to the actual linear instability.

\end{document}